\documentclass[aps,prc,superscriptaddress,showpacs,amsfonts,twocolumn,floatfix]{revtex4}
\usepackage{graphics}
\usepackage{latexsym}
\usepackage{subfigure,wrapfig}
\usepackage{epsfig,color,rotating,amsmath,delarray,array}
\usepackage{makeidx,pifont,float,amssymb}
\definecolor{gray01}{gray}{0.9}
\definecolor{gray02}{gray}{0.8}
\definecolor{gray03}{gray}{0.7}
\definecolor{gray04}{gray}{0.6}
\definecolor{gray05}{gray}{0.5}
\definecolor{gray06}{gray}{0.4}
\definecolor{gray07}{gray}{0.3}
\definecolor{gray08}{gray}{0.2}
\definecolor{gray09}{gray}{0.1}

\begin{document}

\title{Photoproduction of Neutral Pions off Protons}

\newcommand*{\HISKP}{Helmholtz-Institut f\"ur Strahlen- und Kernphysik, Universit\"at Bonn, D-53115 Bonn, Germany}
\newcommand*{\GATCHINA}{Petersburg Nuclear Physics Institute, RU-188350 Gatchina, Russia}
\newcommand*{\PI}{Physikalisches Institut, Universit\"at Bonn, D-53115 Bonn, Germany}
\newcommand*{\KVI}{KVI, 9747 AA Groningen, The Netherlands}
\newcommand*{\FSU}{Department of Physics, Florida State University, Tallahassee, Florida 32306, USA}
\newcommand*{\GIESSEN}{II. Physikalisches Institut, Universit\"at Gie\ss en, D-35392 Gie\ss en, Germany}
\newcommand*{\BASEL}{Physikalisches Institut, Universit\"at Basel, CH-4056 Basel, Switzerland}
\newcommand*{\JUELICH}{Institut f\"ur Kernphysik, Forschungszentrum J\"ulich, D-52428 J\"ulich, Germany}

\author{V.~Crede} \affiliation{\FSU}
\author{N.~Sparks} \affiliation{\FSU}
\author{A.~Wilson} \affiliation{\FSU}
\author{A.V.~Anisovich} \affiliation{\HISKP}\affiliation{\GATCHINA}
\author{J.C.S.~Bacelar} \affiliation{\KVI}
\author{R.~Bantes} \affiliation{\PI}
\author{O.~Bartholomy} \affiliation{\HISKP}
\author{D.~Bayadilov} \affiliation{\HISKP}\affiliation{\GATCHINA}
\author{R.~Beck} \affiliation{\HISKP}
\author{Y.A.~Beloglazov} \affiliation{\GATCHINA}
\author{R.~Castelijns} \affiliation{\KVI}
\author{H.~Dutz} \affiliation{\PI}
%\author{A.~Ehmanns} \affiliation{\HISKP}
\author{D.~Elsner} \affiliation{\PI}
%\author{K.~Essig} \affiliation{\HISKP}
\author{R.~Ewald} \affiliation{\PI}
%\author{I.~Fabry} \affiliation{\HISKP}
\author{F.~Frommberger} \affiliation{\PI}
%\author{M.~Fuchs} \affiliation{\HISKP}
\author{Chr.~Funke} \affiliation{\HISKP}
%\author{R.~Gothe} \affiliation{\PI}
\author{R.~Gregor} \affiliation{\GIESSEN}
\author{A.~Gridnev} \affiliation{\GATCHINA}
\author{E.~Gutz} \affiliation{\HISKP}
\author{W.~Hillert} \affiliation{\PI}
%\author{St.~H\"offgen} \affiliation{\PI}
\author{P.~Hoffmeister} \affiliation{\HISKP}
%\author{I.~Horn} \affiliation{\HISKP}
\author{I.~Jaegle} \affiliation{\BASEL}
\author{J.~Junkersfeld} \affiliation{\HISKP}
\author{H.~Kalinowsky} \affiliation{\HISKP}
\author{S.~Kammer} \affiliation{\PI}
\author{Frank~Klein} \affiliation{\PI}
\author{Friedrich~Klein} \affiliation{\PI}
\author{E.~Klempt} \affiliation{\HISKP}
%\author{M.~Konrad} \affiliation{\PI}
\author{M.~Kotulla} \affiliation{\BASEL}\affiliation{\GIESSEN}
\author{B.~Krusche} \affiliation{\BASEL}
%\author{J.~Langheinrich} \affiliation{\PI}
\author{H.~L\"ohner} \affiliation{\KVI}
\author{I.V.~Lopatin} \affiliation{\GATCHINA}
%\author{J.~Lotz} \affiliation{\HISKP}
\author{S.~Lugert} \affiliation{\GIESSEN}
\author{D.~Menze} \affiliation{\PI}
\author{T.~Mertens} \affiliation{\BASEL}
\author{J.G.~Messchendorp} \affiliation{\KVI}\affiliation{\GIESSEN}
\author{V.~Metag} \affiliation{\GIESSEN}
\author{M.~Nanova} \affiliation{\GIESSEN}
\author{V.A.~Nikonov} \affiliation{\HISKP}\affiliation{\GATCHINA}
\author{D.~Novinski} \affiliation{\GATCHINA}
\author{R. Novotny} \affiliation{\GIESSEN}
\author{M.~Ostrick} \affiliation{\PI}
\author{L.M. Pant} \affiliation{\GIESSEN}
\author{H.~van~Pee} \affiliation{\HISKP}
\author{M.~Pfeiffer} \affiliation{\GIESSEN}
\author{A.~Roy} \affiliation{\GIESSEN}
\author{A.V.~Sarantsev} \affiliation{\HISKP}\affiliation{\GATCHINA}
\author{S.~Schadmand} \affiliation{\JUELICH}
\author{C.~Schmidt} \affiliation{\HISKP}
\author{H.~Schmieden} \affiliation{\PI}
\author{B.~Schoch} \affiliation{\PI}
\author{S.~Shende} \affiliation{\KVI}
\author{V.~Sokhoyan} \affiliation{\HISKP}
\author{A.~S{\"u}le} \affiliation{\PI}
\author{V.V.~Sumachev} \affiliation{\GATCHINA}
\author{T.~Szczepanek} \affiliation{\HISKP}
\author{U.~Thoma} \affiliation{\HISKP}
\author{D.~Trnka} \affiliation{\GIESSEN}
\author{R.~Varma} \affiliation{\GIESSEN}
\author{D.~Walther} \affiliation{\HISKP}\affiliation{\PI}
%\author{Ch.~Weinheimer} \affiliation{\HISKP}
\author{Ch.~Wendel} \affiliation{\HISKP}
\collaboration{The CBELSA/TAPS Collaboration} \noaffiliation

\begin{abstract}
Photoproduction of neutral pions has been studied with the CBELSA/TAPS detector in
the reaction $\gamma p\to p\pi^0$ for photon energies between 0.85 and 2.50~GeV. The 
$\pi^0$~mesons are observed in their dominant neutral decay mode: $\pi^0\to\gamma\gamma$.
For the first time, the differential cross sections cover the very forward region,
$\theta_{\rm c.m.}<60^\circ$. A partial-wave analysis of these data within the Bonn-Gatchina framework 
observes the high-mass resonances $G_{17}$(2190), $D_{13}$(2080), and $D_{15}$(2070).
\end{abstract}

\date{Received: \today / Revised version:}


\maketitle

\section{\label{Section:Introduction}Introduction}
Spectroscopy has long provided essential information for scientists trying to understand
the nature of particles. The spectroscopy of the hydrogen atom revealed the quantum nature 
of the interactions between particles at small distance scales and led to the development 
of quantum electrodynamics (QED). Exploration of the excited states of hadrons opens a door 
into the nature of nonperturbative quantum chromodynamics (QCD). 
%By mapping out the excitation spectra of hadrons, information is gained about the effective number of degrees of 
%freedom and the effective forces that bind the constituents within hadrons. 
By mapping out the excitation spectra of hadrons, information is gained about their effective degrees of 
freedom and the effective forces that lead to confinement. 
A both particularly challenging and interesting task is to understand the spectra of baryons. Although 
symmetric quark models, describing the baryon with three constituent quark degrees of freedom, 
are good at predicting many properties for many of the low-lying baryons (especially the ground states), 
%many properties of the ground-state baryons and the corresponding first excitations in each partial wave, 
there is a large disagreement between the number of excited 
states predicted by these models and the number that has been observed at and, in particular, 
above masses of about 1.8~GeV/$c^2$. Attempts at finding a physical mechanism that can account 
for this discrepancy have not thus far yielded a satisfactory answer, and the higher mass regions 
of the excitation spectra remain largely unexplored: Most known baryon resonances have masses 
smaller than 2~GeV/$c^2$ and were discovered in elastic $\pi N$ scattering experiments. Many 
of the {\it missing} or hitherto unobserved baryon resonances are predicted to couple only 
weakly to the $\pi N$~channel but exhibit large couplings to $\gamma N$. In recent years, 
many different laboratories worldwide (Jefferson Laboratory, ELSA, MAMI, GRAAL, SPring-8, etc.) have 
measured differential cross sections and polarization observables for a large variety of final 
states in electromagnetically induced reactions. A good review is given in~\cite{Klempt:2009pi}. 
%Both unpolarized cross sections and polarization observables are needed to maximally constrain 
%and uniquely determine the scattering amplitude for a given process. 

Although many of the {\it missing} baryon resonances may have a small coupling   
to $\pi N$, it is still important to study pion photoproduction. Doing so may confirm or reject 
%resonances seen in elastic pion-nucleon scattering, and it maywhile at the same time providing evidence for unexplored
resonances seen in elastic pion-nucleon scattering, and it could still provide evidence for unexplored
resonances. New resonances found in reactions such as
%$\gamma N\to\pi N$ are expected to have masses smaller than 2.5~GeV/$c^2$, while higher-mass
$\gamma N\to\pi N$ are expected to have masses larger than about 1.8~GeV/$c^2$, although the higher-mass
resonance contributions are expected to be more important in double-meson photoproduction, by 
%sequentially decaying via emission of a $\pi$ or $\eta$ meson and populating intermediate 
sequentially decaying via emission of a $\pi$ or $\eta$ meson and populating intermediate 
states.
%, based, experimentally, on the relatively small resonant single-pion photoproduction cross section at higher masses.

\begin{figure*}[ht]
  %\vspace{0.8cm}
  %\epsfig{file=fig1.eps,width=0.9\textwidth,height=0.39\textwidth}
  \epsfig{file=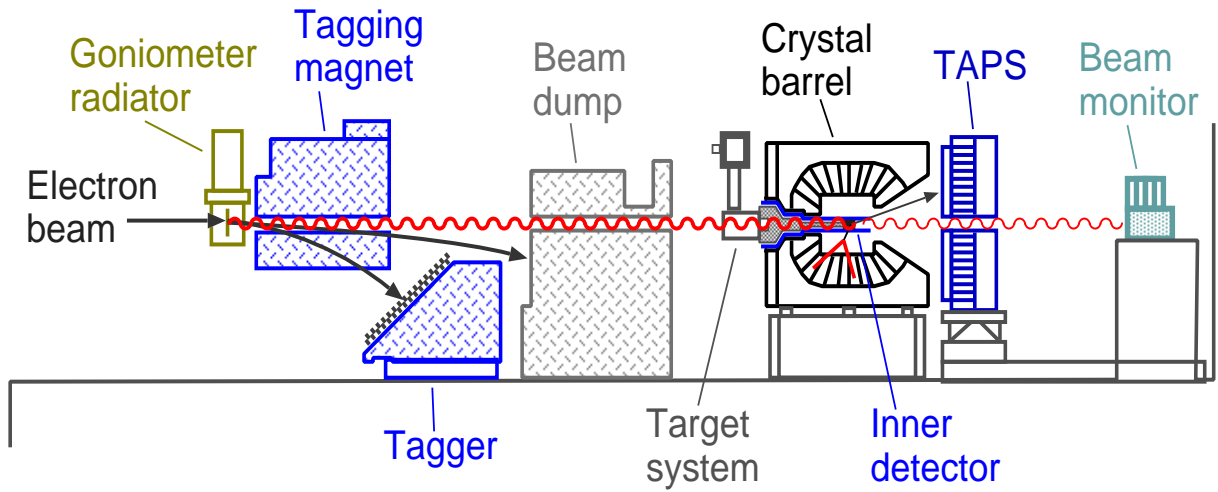,width=0.9\textwidth,height=0.37\textwidth}
  %\vspace{-0.5cm}
  \caption{\label{Figure:Experiment} (Color online) Cross section of
    the CBELSA/TAPS setup (not to scale). 
     The electron beam delivered by the accelerator ELSA enters from the left side
     and hits the copper radiator (for the data presented here) in front of the tagger magnet.}
\end{figure*}

Excited baryons are broad and overlapping, making it difficult to extract their properties by studying isolated
reaction channels and cross sections alone. 
%Both unpolarized cross sections and polarization observables are 
%needed to maximally constrain and uniquely determine the scattering amplitude for a given process; 
Polarization observables can provide access to small resonance contributions to a final state, so that 
{\it missing} resonances may still be found by
studying these in pion photoproduction. However, both unpolarized cross sections and polarization observables are 
needed to maximally constrain and uniquely determine the scattering amplitude for a given process. Additional
physical constraints are provided by performing an analysis of these observables for multiple reaction channels
simultaneously, and to succeed in its description
of the contributing resonances, the combined analysis of these
data must accurately describe the nonresonant behavior of the
observables. Previous experiments have shown that $t$-channel
contributions are important in pion
photoproduction at high energies, where the {\it missing} resonances are expected. Thus further study of the pion
photoproduction cross section, particularly at forward angles, is very interesting. 

%Additional physical constraints are provided by measuring polarization
%observables and also by studying observables for a large number of channels. 
%The pion, the lightest meson, plays a large role in the spectroscopy of baryons
%because many resonances couple to it as a result of the enhanced 
%phase space. Studying the pion-nucleon final state provides access to both $N^*$ and $\Delta$ resonances,
%so it is vital to understand its reactions well. 
% the pion-nucleon final state is vital because isospin conservation allows 
%the strong decay of both $N$ and $\Delta$ excitations to it.
%$N$($I=1/2$) and $\Delta$($I=3/2$)
%Therefore, it is vital to understand reactions that have a pion and nucleon in the 
%final state.
% Pion-nucleon scattering has dominated our knowledge of the excited nucleon states, 
%and it is imperative to strengthen this experimental evidence as well as search for 
%new resonances in pion photoproduction. 

Recent measurements of differential cross sections in pion photoproduction have made 
significant contributions to the world database, but these measurements have limited 
coverage at very forward center-of-mass polar angles of the pion, $\theta_{\rm c.m.}$.
%in that they do not cover the very forward direction. 
%Limited angular coverage renders it difficult for an 
%amplitude analysis to extract both resonant and nonresonant contributions. 
The recent CB-ELSA~\cite{vanPee:2007tw,Bartholomy:2004uz} and CLAS data~\cite{Dugger:2007bt} on 
$\gamma p\to p\pi^0$ cover angles down to approximately $30^{\circ}$ and 
$40^{\circ}$ in $\theta_{\rm c.m.}$, respectively. They show discrepancies, where the respective 
differential cross sections are observed to increase at different rates as one moves to more 
forward angles for the incoming photon energy range 1300 to 2000 MeV. Our new CBELSA/TAPS 
data, presented here, extend the coverage of the forward region and will help understand the 
behavior of the $\pi^0$~photoproduction cross section in this kinematic region.

In this paper, we present differential cross sections for the reaction:
\begin{align}
\gamma p\to p\pi^0,&\quad\text{ where }\pi^0\to 2\gamma.\label{Reaction}
\end{align}
\noindent
We studied, in particular, the forward region; a similar study was
published for the beam asymmetries
$\Sigma$ in the same reaction~\cite{Sparks:2010vb}.

The paper has the following structure. Section~\ref{Section:PreviousResults} summarizes 
recent results on $\pi^0$~photoproduction. A brief introduction to the CBELSA/TAPS experimental 
setup is given in Section~\ref{Section:ExperimentalSetup}. The data reconstruction and event
selection are discussed in Section~\ref{Section:DataAnalysis}, and the extraction of cross 
sections is described in Section~\ref{Section:Extraction}. Experimental results are finally 
presented in Section~\ref{Section:Results}, and results of a
partial-wave analysis are discussed in Section~\ref{Section:PWA}.

\section{\label{Section:PreviousResults}Previous Results}
Cross section data for $\pi^0$~photoproduction were obtained and studied at many 
different laboratories over a wide kinematic range~\cite{Brefeld:1975dv,Althoff:1979mc,
Yoshioka:1980vu,Bergstrom:1997jc,Beck:1990da,Beck:1997ew,Fuchs:1996ja,Krusche:1999tv,
Schmidt:2001vg,Blanpied:2001ae,Ahrens:2002gu,vanPee:2007tw,Bartholomy:2004uz,Bartalini:2005wx,Dugger:2007bt,
Sumihama:2007qa}. A review of the main data sets published before 2005 and a corresponding 
comparison of their coverage in energy and solid angle can be found in~\cite{vanPee:2007tw}.
The new data presented in this paper extend previous data by covering the very forward polar angles
with an electromagnetic calorimeter, TAPS.

In 2005, the CB-ELSA collaboration at Bonn, Germany, presented high-statistics results on the 
photoproduction of $\pi^0$~mesons using the Crystal Barrel detector for incident photon energies 
from 300 to 3000~MeV~\cite{Bartholomy:2004uz}. Above energies of 1300~MeV, the data cover an 
angular range from about $30^\circ$ to $140^\circ$, which corresponds to about $-0.75 < {\rm 
cos}\,\theta_{\rm c.m.} < 0.85$.

The GRenoble Anneau Accelerateur Laser (GRAAL) collaboration at ESRF in Grenoble, France, measured 
differential cross sections over a wide angular range from 550 to 1500~MeV~\cite{Bartalini:2005wx}, 
though still limited to ${\rm cos}\,\theta_{\rm c.m.}~<~0.7$ or approximately $\theta_{\rm c.m.} > 45^\circ$. 
At the GRAAL facility, Compton 
backscattering of low-energy photons off ultrarelativistic electrons reached almost 100\,\% beam 
polarization at the Compton edge. 

More recently, the LEPS collaboration at SPring-8 in Hyogo, Japan, published results for higher 
photon energies between $E_\gamma$~=~1500 and 2400~MeV~\cite{Sumihama:2007qa}. For the first time, 
the data cover the $\pi^0$~backward angles between $-1 < {\rm cos}\,\theta_{\rm c.m.} < -0.6$ or about
$125^\circ < \theta_{\rm c.m.} < 180^\circ$.
Backward-Compton scattering was applied using Ar-ion laser photons with a 351-nm wavelength.

At Jefferson Lab, Newport News, VA, differential cross sections for the reaction were 
measured using the CEBAF Large Acceptance Spectrometer (CLAS) and a tagged photon beam with 
energies from 0.675 to 2.875~GeV~\cite{Dugger:2007bt}. The cross-section data cover an angular 
range from $\theta_{\rm c.m.}\approx 40^\circ$ to $150^\circ$, corresponding to approximately 
$-0.85 < {\rm cos}\,\theta_{\rm c.m.} < 0.75$.

\section{\label{Section:ExperimentalSetup}Experimental Setup}

\begin{figure}[t]
  \begin{center}
    \epsfig{file=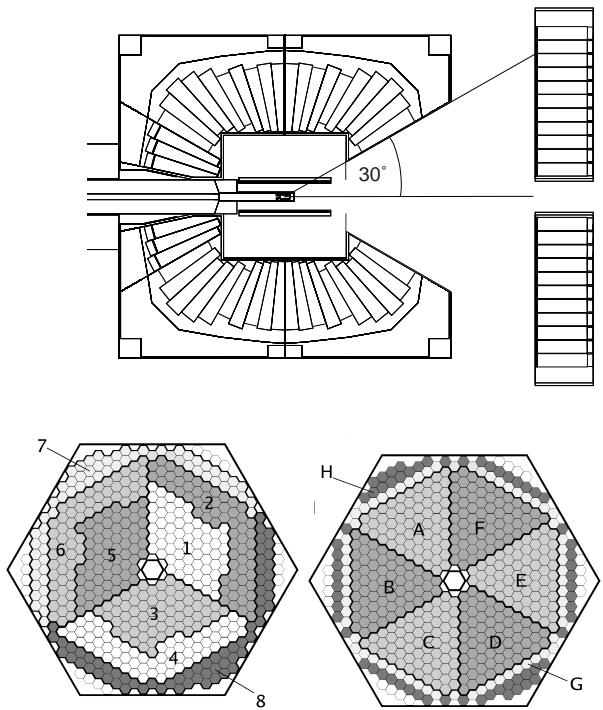,width=0.5\textwidth}
  \end{center}
  \caption{\label{Figure:CB-Luzy-H2} (Top) Schematic drawing of the liquid-hydrogen 
    target, scintillating-fiber detector, Crystal Barrel and TAPS calorimeters. 
    (Bottom) Front view of TAPS: The left side shows the logical segmentation for the 
    LED-low trigger; the right side shows the logical segmentation for the LED-high trigger 
    (see text for more details).}
\end{figure}

The experiment was conducted using the Crystal Barrel~\cite{Aker:1992} and TAPS
\cite{Gabler:1994ay,Novotny:1991ht} electromagnetic calorimeters at the electron stretcher 
accelerator ELSA~\cite{Hillert:2006yb}, which is located at the University of Bonn in 
Germany. Figure~\ref{Figure:Experiment} contains a schematic of the experimental setup.

\begin{figure*}[t]
  \epsfig{file=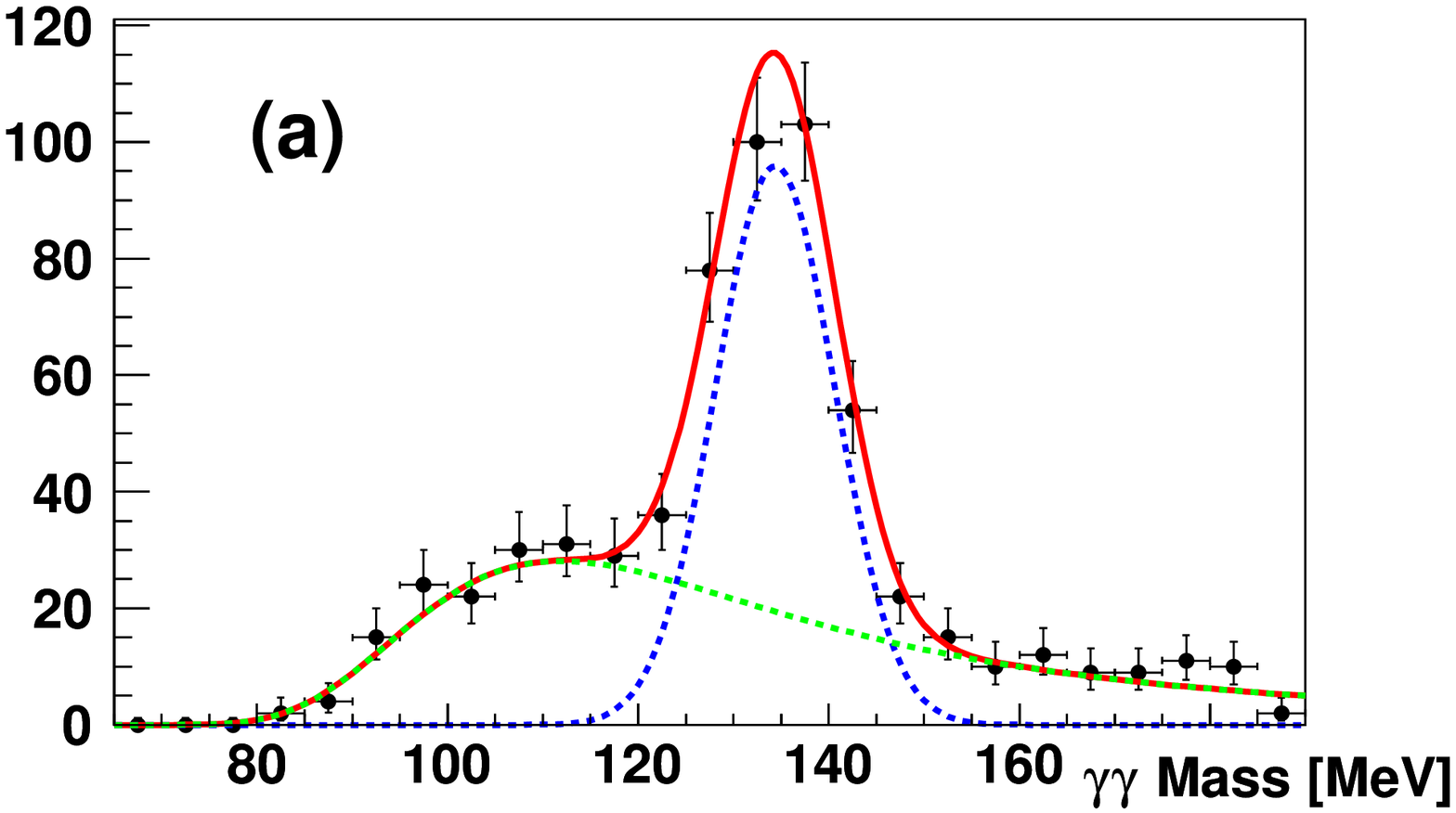,width=.325\textwidth} \hfill
  \epsfig{file=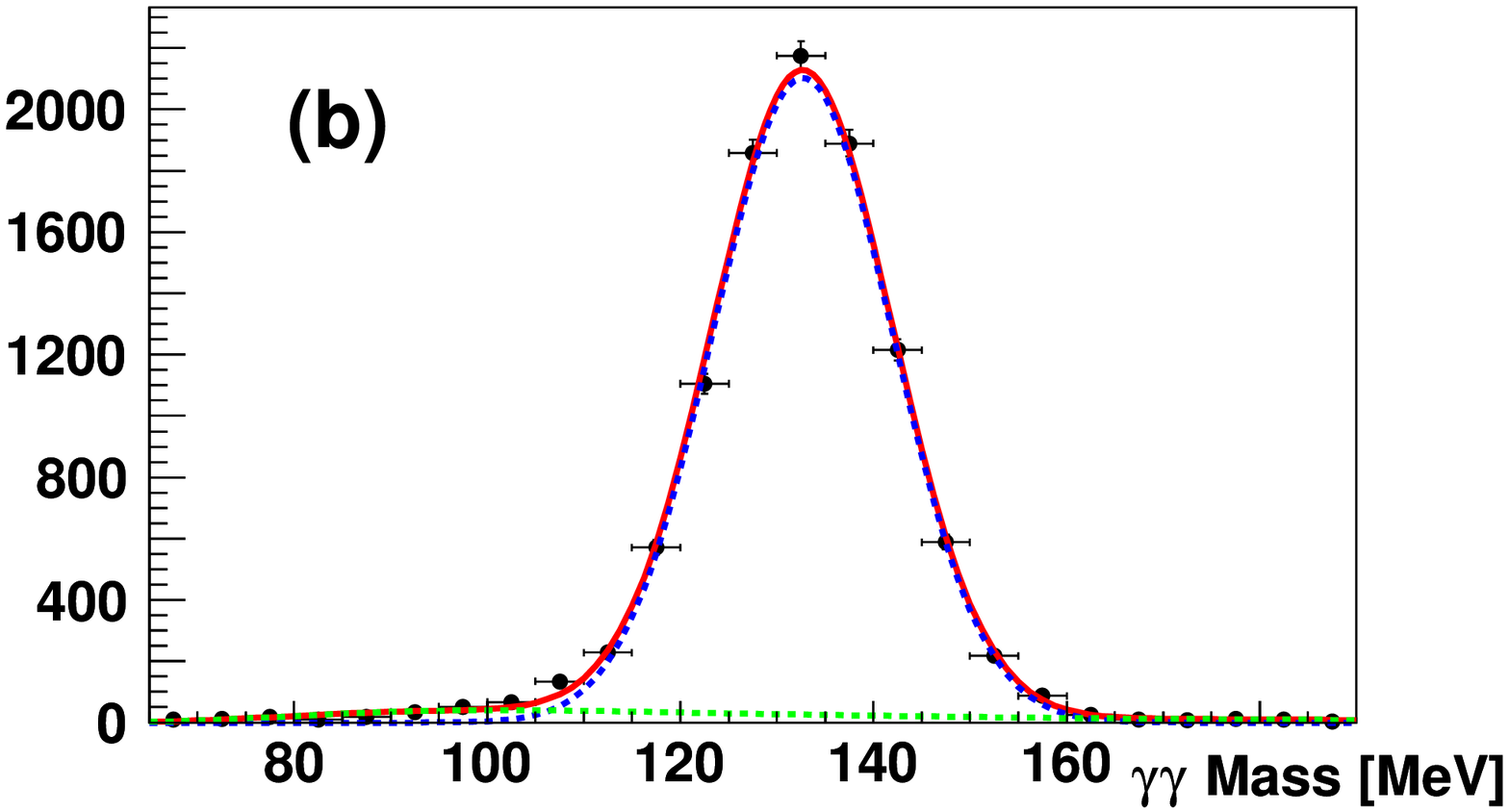,width=.325\textwidth} \hfill
  \epsfig{file=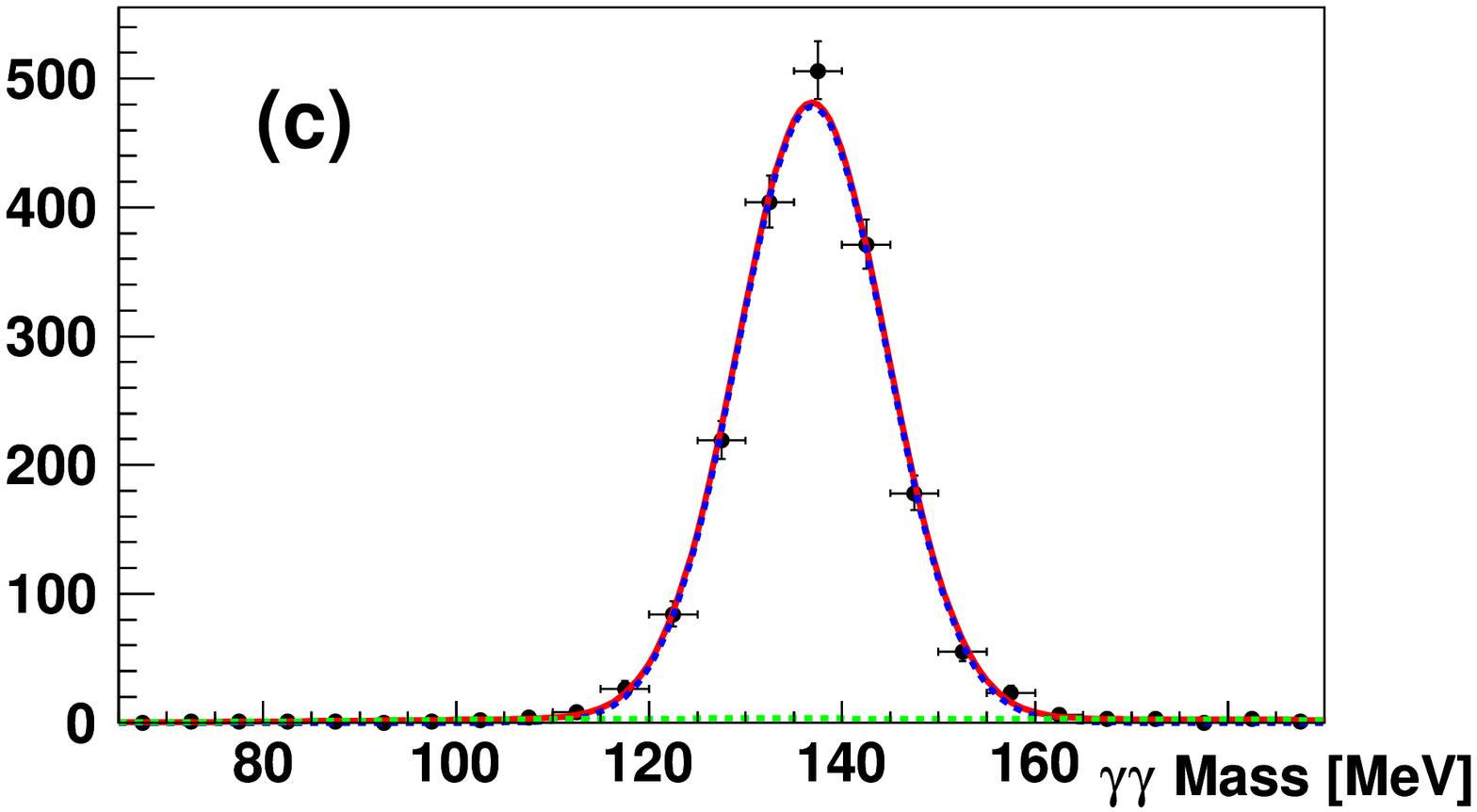,width=.325\textwidth}
  \caption{\label{ggMass}(Color online) Invariant $\gamma\gamma$ mass spectra for the 
    reaction $\gamma p\to p\gamma\gamma$ for $E_\gamma\in [875,900]$~MeV and for (a) 
    $\theta_{\rm c.m.}\in [0^\circ,10^\circ]$, (b) $\theta_{\rm c.m.}\in [40^\circ,50^\circ]$, and
    (c) $\theta_{\rm c.m.}\in [160^\circ,170^\circ]$. CL cuts were applied at 
    $10^{-2}$. The red solid line denotes the signal Gaussian, the green dotted line the
    Landau distribution for the background description, and the blue dotted line shows the background 
    subtracted distribution. The background is essentially negligible for $\theta_{\rm c.m.}
    > 50^\circ$.}
\end{figure*}

A 3.175-GeV electron beam was extracted (by slow resonant extraction) from ELSA. The 
electrons were passed through a thin copper radiator, which had a thickness of $(3/1000)~X_{R}$ 
(radiation length), where a fraction of them underwent bremsstrahlung. These electrons 
were deflected by a dipole magnet into the tagger detector system (tagger), which consisted 
of 480~scintillating fibers above 14~scintillation counters in a partly overlapping 
configuration. Good knowledge of the magnetic field and the position of the hit in the 
tagger allowed the momentum and energy of each electron to be determined. The tagged-photon 
energy is given by the difference between the incoming electron energy and the electron 
energy after bremsstrahlung. Electrons that did not undergo bremsstrahlung were deflected 
at small angles into a beam dump located behind the tagger. Tagged photons had energies 
ranging from 0.5 to 2.9~GeV; the higher energy part of this bremsstrahlung spectrum had 
an energy resolution of about 2~MeV, while the lower part had an energy resolution of 
approximately 25~MeV.

The energy calibration of the photon tagging system was determined in
simulations using the known positions of the scintillating fibers and
the measured magnetic field map, and approximated by a polynomial. 
Measurements with a known (lower-energy) ELSA beam allowed for a
cross-check of the calibration; this involved guiding a low-current
beam directly into the tagger, while holding the magnetic field
constant. Corrections to the original polynomial were determined from
this procedure.

Tagged photons continued along the beam line toward a liquid-hydrogen target, which was 5~cm 
in length and 3~cm in diameter, located at the center of the Crystal Barrel (CB) calorimeter 
(see Fig.~\ref{Figure:Experiment}). Some of these photons interacted with protons within the 
target to produce a proton and multiple photons in the final state, the photons resulting from 
the decay of neutral mesons. Final-state photons were detected by the two electromagnetic calorimeters 
(CB and TAPS), which together almost covered the full $4\pi$~solid
angle ($>$ 98\,\% coverage). 
The CB detector in its CBELSA/TAPS configuration of 2002/2003 consisted of 1290 CsI(Tl) crystals, 
which had a trapezoidal shape and were oriented toward the target, with a thickness of $16~X_{R}$, 
and covered the polar angles from 30$^{\circ}$ to 168$^{\circ}$, while TAPS consisted of 528 
hexagonal BaF$_2$ crystals with a length of approximately $12~X_{R}$, and covered the polar 
angles from 5$^{\circ}$ to 30$^{\circ}$; both covered the full azimuthal circle. TAPS was 
configured as a hexagonal wall and served as the forward endcap of the CB. The CB originally
covered polar angles down to 12$^{\circ}$ and was made of 26 rings~\cite{Aker:1992}. Downstream rings 
$11-13$ of the CB were removed to allow for a tight fit with TAPS (see top of Fig.~\ref{Figure:CB-Luzy-H2}). 
Forward-going protons were 
detected by plastic scintillators (5~mm thick) located in front of each TAPS module; the other 
protons were detected by a three-layer scintillating fiber detector, which closely surrounded the 
target~\cite{Suft:2005cq}. 

Photons that did not interact with the target material continued further along the beam line 
through a hole in TAPS and finally arrived at the beam monitor, a
total absorption $\check{\rm C}$erenkov 
counter, which consisted of an array of nine lead glass crystals. The number of photons counted 
here was vital to the determination of the photon flux. 

The first-level trigger was provided by the fast response of the TAPS
modules. A cellular logic called FAst-Cluster Encoder (FACE), which
counted the number of clusters in the barrel, formed the second-level
trigger. The trigger required at least one hit in TAPS above a high-energy
threshold (LED-high) and at least one FACE cluster, or it
required at least two hits in TAPS in different segments above a low-energy 
threshold (LED-low). The logical LED-low and LED-high
segmentation of the trigger is shown in Fig.~\ref{Figure:CB-Luzy-H2}. 
Topologically, reaction~(\ref{Reaction}) events satisfying the first
condition have either a fast proton in TAPS (and a backward-going pion
in the CB) or one high-energy photon in TAPS and one photon in the CB 
(from a forward-going pion), while those that satisfy the second
condition have a pion in TAPS, together leaving a hole in the pion
acceptance at intermediate angles.

\section{\label{Section:DataAnalysis}Preparation of Final State}
The data presented here were recorded from October 2002 until November 2002 in two run periods
at ELSA beam energies of 3.175~GeV. For this analysis, only the photon
energy range up to 2.55~GeV was 
used since the tagger consisted of scintillating fibers covering energies up to this energy and 
providing timing information, while simple hit/miss wires covered photon energies above 2.55~GeV. 
These data were also used for previously published analyses on a variety of different final 
states~\cite{Castelijns:2007qt,Nanova:2008kr,Crede:2009zz}. 

This section discusses the event reconstruction and selection of the $\pi^0$~photoproduction
channel: reaction~(\ref{Reaction}). A total number of $2.21\times 10^6~\pi^0$~events was observed, 
covering invariant masses from 1570 to 2360~MeV/$c^2$. 

\subsection{\label{Subsection:EventReconstruction}Event Reconstruction}
The event selection was carried out in close analogy to the study of the $\eta$~photoproduction 
channel. Further details on the event reconstruction and the use of kinematic fitting in this
 analysis can be found in~\cite{Crede:2009zz}. 

Events were analyzed for which only the two final-state photons (from the decay of the pion, $\pi^0
\to\gamma\gamma$) or all final-state particles ($\gamma\gamma + p$) were detected. The protons 
were identified by using the photon-veto BaF$_2$ detectors of TAPS or the inner scintillating
fiber detector inside the CB. A CB cluster is assigned to a charged particle if the
trajectory from the target center to the barrel hit forms an angle of less than $20^\circ$ with
a trajectory from the target center to a hit in the inner detector. The resolution for proton 
clusters in the calorimeters is worse than that for photon clusters because, on average, proton 
clusters are much smaller than photon clusters. Proton identification was only used to separate 
protons from photons, but the proton four-vector was determined from the event kinematics through 
missing-proton kinematic fitting.

Time accidental background was reduced by demanding a prompt coincidence between an electron in 
the tagger and a particle in TAPS; if this particle was a photon, it required a time difference 
within $\pm 3$~ns, while for a proton a time difference within -5 to +15~ns was required.
Two to three tagger photons, at most, survived the initial timing cut. 
%A missing-proton kinematic fit also allowed the identification of the initial-state photon in the event. 
A missing-proton kinematic fit was used to separate the best initial-state photon from the remaining candidates in the event. 
This 1C kinematic fit to the hypothesis $\gamma p\to p_{\rm \,miss}\,\gamma\gamma$, which only required four-momentum 
conservation, was performed for each initial photon candidate of an event passing the initial timing 
cut, and the candidate with the largest confidence level (CL) or $\chi^2$~probability was selected. 
%Two to three tagger photons at most survived the initial timing cut. 
Finally, only events with a CL $> 1\,\%$ were accepted for further analysis. No mass constraints were used in the 
kinematic fitting.

In the final step of the event reconstruction, we removed events that are not $p(\pi^0
\to\gamma\gamma)$ final states (background) from the event sample. This was done by interpolating 
the shape of the background in the signal region based on the distribution of events outside the 
signal region. The data were binned, and maximum likelihood fits of a signal plus background 
hypothesis to the data were performed. A Gaussian line shape was used for the signal, and a Landau
distribution was used for the background. 
Typical $\gamma\gamma$~mass distributions are shown in 
Fig.~\ref{ggMass} for $E_\gamma\in [875,900]$~MeV. The size of the background at very forward angles 
is substantial [Fig.~\ref{ggMass}\,(a)]; it is much smaller, on average, in the backward bins at less 
than 4\,\% [Fig.~\ref{ggMass}\,(c)]. 
Mass resolutions (standard deviations of Gaussian signals) extracted from the data agree within 1 MeV/c$^2$ 
with those from Monte Carlo simulations.
%Monte Carlo simulations gave mass resolutions that were in good agreement with the data. 

\subsection{\label{Section:MonteCarloSimulations}Monte Carlo Simulations}
The performance of the detector was simulated in GEANT3-based Monte Carlo studies. We used 
a program package that has been built upon a program developed for the CB-ELSA experiment. 
The Monte Carlo program accurately reproduces the response of the TAPS and CB
crystals to photons.

The acceptance for reaction (\ref{Reaction}) was determined by
simulating events that were evenly distributed over the available phase space. 
The tagging and timing of initial-state photons were not simulated.
The Monte Carlo events were analyzed by using 
the same reconstruction criteria as were also applied to the (real) measured data.
%if the criteria were part of the simulation.
The same 1C-hypothesis was tested in the kinematic fits, and events
were selected with the same 
CL cuts. The acceptance is defined as the ratio of the number of reconstructed to 
generated Monte Carlo events,
\begin{equation}
  A_{\gamma p\,\to\,p\,X}=\frac{N_{\rm rec,MC}}{N_{\rm gen,MC}}
     \qquad (X=\pi^0\,)\,.
\end{equation}
\noindent
In the analysis presented here, we have applied an acceptance cut of at least 5\,\% on ($E_\gamma,
\,{\rm cos}\,\theta_{\rm c.m.}\,$) bins, as well as an acceptance cut of at least 1\,\% on ($E_\gamma,\,
\theta_{\rm c.m.}\,$) bins, and removed the other data points from the
analysis. Different values of the acceptance cut were used for the two
binnings to provide comparable statistics over a large angular range.

\section{Determination of Cross Sections\label{Section:Extraction}}
The differential cross sections for this analysis have been determined according to
\begin{align}
\frac{\rm d\sigma}{\rm d\Omega}&=\frac{N_{\,\pi^0\,\to\,\gamma\gamma}}{A_{\,\pi^0\,\to\,
\gamma\gamma}}\medspace\frac{1}{N_{\gamma}\,\rho_{\,\rm t}}
\medspace\frac{1}{\Delta\Omega}\medspace\frac{1}
{\frac{\Gamma_{\pi^0\,\to\,\gamma\gamma}}{\Gamma_{\rm\, total}}}\,,\label{gldsigdom}
%\vspace{-10mm}
\end{align}
where

\begin{table}[H]
\begin{tabular}{rl}
$\rho_{\,\rm t}$\,: & target area density\\
N$_{\,\pi^0\,\to\,\gamma\gamma}$\,: & number of reconstructed data events\\
 & in an ($E_\gamma$,\,cos\,$\theta_{\,\rm c.m.}$) or ($E_\gamma$,\,$\theta_{\,\rm c.m.}$) bin\\
$N_\gamma$\,:& number of beam photons in an ($E_\gamma$) bin\\
A$_{\,\pi^0\to\gamma\gamma}$\,: & acceptance in an ($E_\gamma$,\,cos\,$\theta_{\,\rm c.m.}$) or 
                  ($E_\gamma$,\,$\theta_{\,\rm c.m.}$) bin\\[0.6ex]
$\Delta\Omega$\,: & solid-angle interval $\Delta\Omega= 2\pi\Delta \rm
cos\,(\theta_{\,\rm c.m.})$\\
$\frac{\Gamma_{\,\pi^0\to\gamma\gamma}}{\Gamma_{\rm\, total}}$\,: & decay branching fraction.
\end{tabular}
\end{table}

The target area density, that is, the number of atoms in the target material per cross-sectional 
area (orthogonal to the photon beam), is given by
\begin{equation}
\rho_{\rm t} = 2\,\frac{\rho({\rm H}_2) N_A L}{M_{\rm mol}({\rm H}_2)} = 2.231\cdot 10^{-7}
\mu {\rm b}^{-1}\,,
\end{equation}
where $\rho({\rm H}_2) = 0.0708$ g/cm$^3$ is the density, $M_{\rm mol} = 2.01588$~g/mol is the 
molar mass of liquid H$_2$, $N_A = 6.022\cdot 10^{23}$ mol$^{-1}$ is the Avogadro number, and 
$L = 5.275$ cm is the length of the target cell. The factor of two accounts for the molecular 
composition of hydrogen (H$_2$).

The number of events in an ($E_\gamma$,\,cos\,$\theta_{\rm c.m.}$) or ($E_\gamma$,\,
$\theta_{\rm c.m.}$) bin comprises events with two or three final-state particles (at least 
$2\,\gamma$'s). The proton can be ``missing'', but events with and
without a detected proton 
are treated in the same way in the event reconstruction. For ${\rm cos}\,\theta_{\rm c.m.} 
< 0.0$, the event kinematics requires that the proton is used in the (TAPS) trigger. Thus, 
the detection efficiency of forward-going protons needs to be reasonably well understood. 
This has been shown earlier for CBELSA/TAPS results on $\eta$~photoproduction; more details 
on the proton trigger can be found in~\cite{Crede:2009zz}.

The interval of the solid angle is given by $\Delta\,\Omega=2\pi\Delta\,\cos(\theta_{\rm c.m.})$, 
with $\Delta\,\cos(\theta_{\rm c.m.})$ describing the width of the angular bins. It was 
chosen to be $0.1$ for the $\pi^0$~data presented here. For the representation of the 
differential cross sections in terms of the angle $\theta_{\rm c.m.}$ directly, a bin width 
of $10^\circ$ was chosen. Energy bins were defined by considering statistics and ensuring a 
good comparability with other experiments. A total of 42~bins is presented in energy steps of 
25~MeV for $E_{\gamma}\,\in\,[850,~1300]$~MeV and 50~MeV for $E_{\gamma}\,\in\,[1300,~2500]$~MeV.

The number of observed $\pi^0$~mesons needs to be corrected for unseen decay modes. 
The partial-decay branching fraction used to correct the measured cross sections is taken 
from~\cite{Nakamura:2010zzi}: BR$(\pi^0\to\gamma\gamma) = 0.98823\,\pm\, 0.00034$.

\subsection{Normalization\label{Subsection:Normalization}}
The CBELSA/TAPS tagging hodoscope consisted of 480 scintillating fibers above 14 partially 
overlapping scintillation counters (tagger bars). The photon flux was measured directly in 
the experiment and determined according to
\begin{equation}
N_\gamma\,=\, \alpha \cdot P_\gamma \cdot N\,^{\rm fiber}_{\rm scaler}\,,
\end{equation}

\noindent
where the parameter $\alpha$ accounts for the (fiber)-cluster reconstruction in the tagger, 
which has to be performed in the same way as for real hadronic events. $P_\gamma$, the photon 
definition probability, denotes the probability that a real photon is emitted along the beam 
axis in the tagger and traverses the liquid hydrogen target; it is determined from Tagger-Or-Runs, 
separate data runs utilizing a {\it minimum-bias} trigger. $N\,^{\rm fiber}_{\rm scaler}$ is the 
number of {\it free} hardware counts for the individual fibers corrected for the lifetime of the 
detector. Scalers are recorded in {\it scaler} events, which were accumulated with a 
minimum-bias trigger at a rate of 1\,Hz during regular data taking. This trigger
required only a hit in the tagger and was thus independent of hadronic cross sections. 

\begin{figure*}[pt]
  \epsfig{file=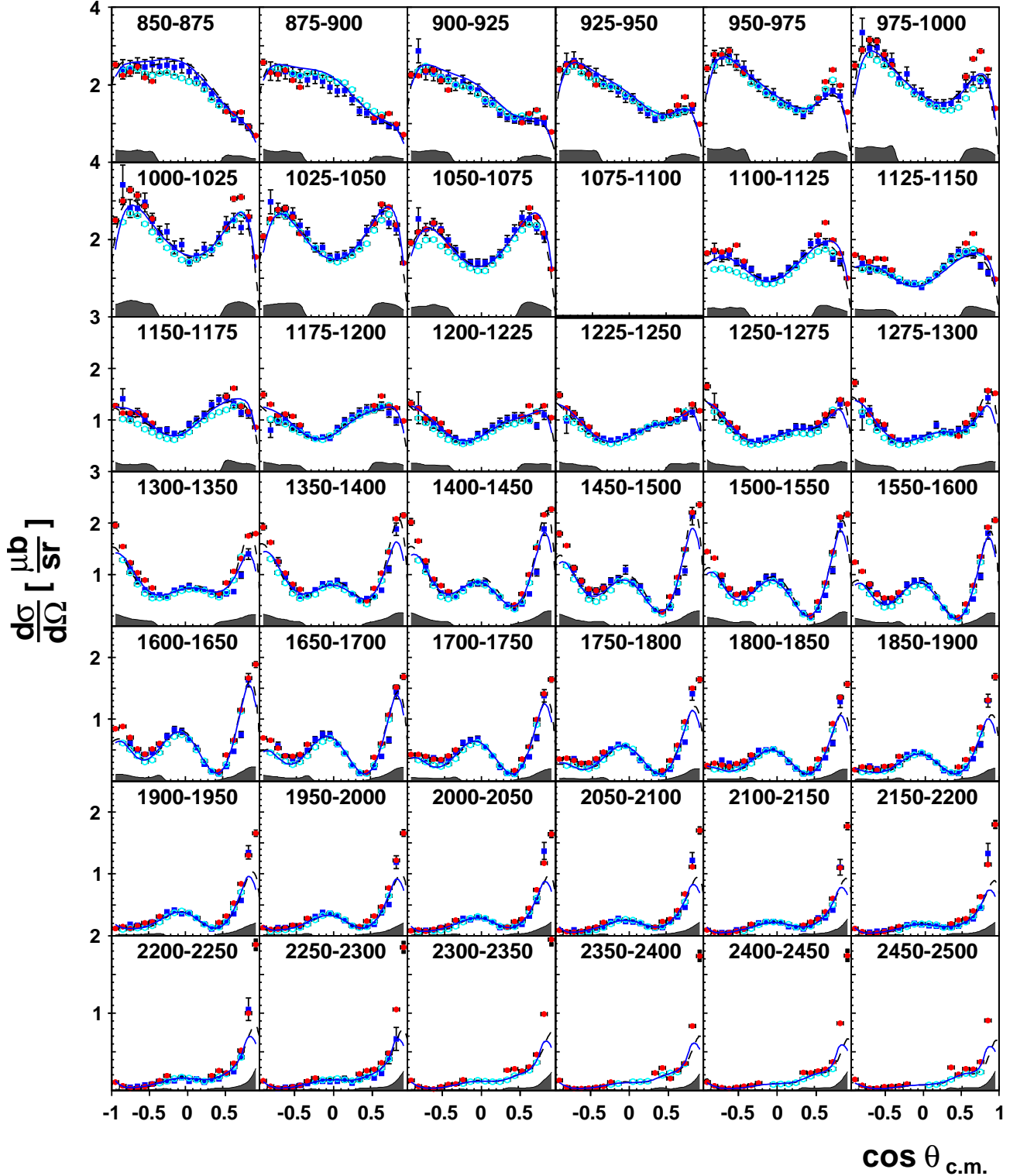,width=1.0\textwidth}
  \vspace{-1.2cm} 
  \caption{\label{Figure:CrossSections}(Color online) Differential cross sections for $\gamma p\to 
    p\pi^0$ from CBELSA/TAPS ({\large\color{red}$\bullet$}) plotted versus cos\,$\theta_{\rm c.m.}$. 
    For comparison, CB-ELSA data~\cite{Bartholomy:2004uz} are represented by 
    ({\tiny\color{blue}$\blacksquare$}) and CLAS data~\cite{Dugger:2007bt} by 
    ({\large\color{cyan}$\circ$}). Note that the CLAS cross sections were extracted for 50-MeV-wide
    energy bins; for $E_\gamma < 1.3$~GeV, the same CLAS results are
    thus shown in the corresponding 25-MeV-wide bins chosen for this
    analysis. The solid blue line shows the solution of the Bonn-Gatchina
    partial-wave analysis~\cite{Anisovich:2010an}, and the dashed black 
    line represents the current SAID solution (SP09)~\cite{Dugger:2009pn}. 
    The data points include statistical errors only; the total
    systematic error is given as error bands at the bottom of each plot.}
\end{figure*}

\begin{figure*}[pt]
  \epsfig{file=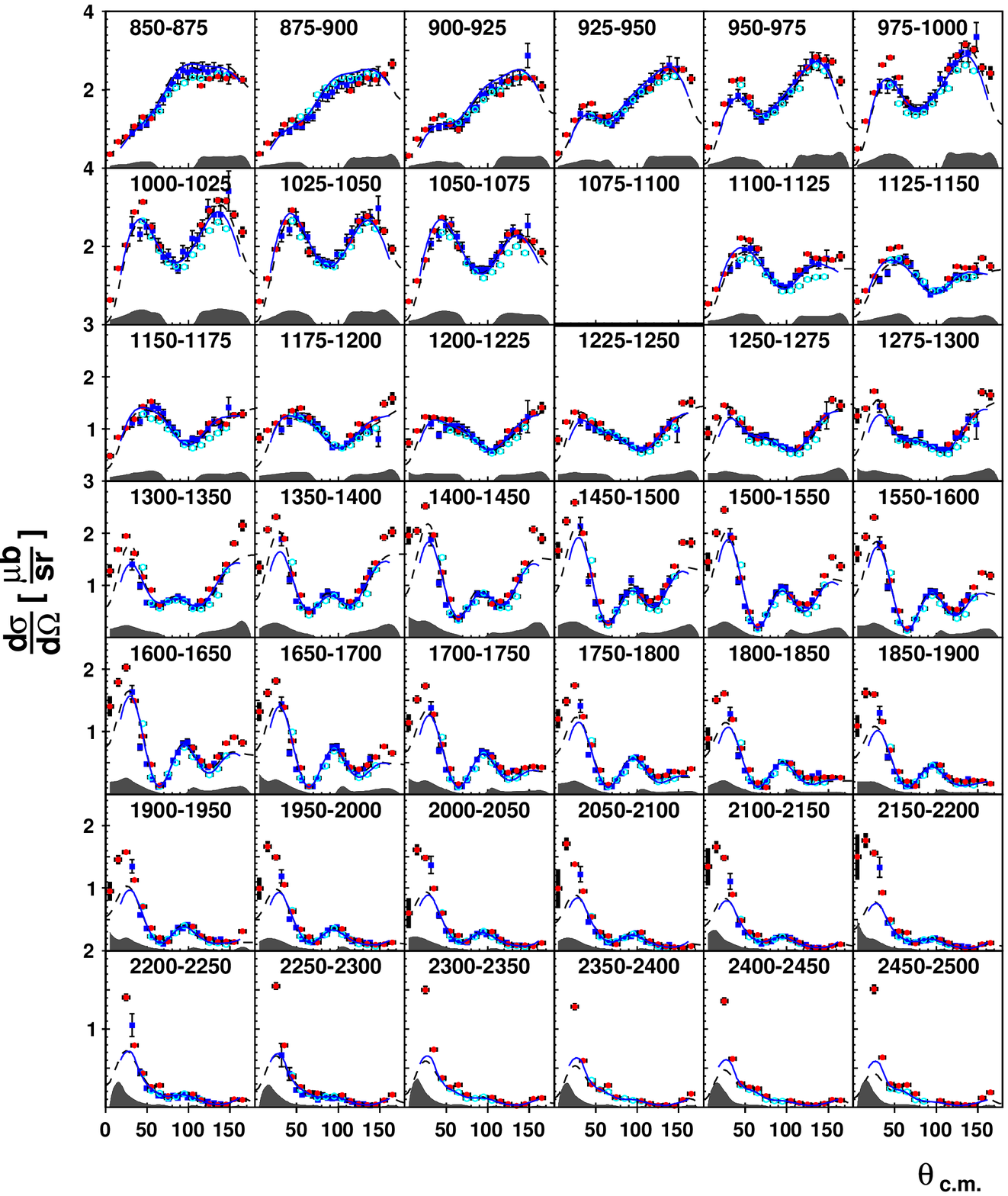,width=1.0\textwidth}
  \vspace{-1.2cm} 
  \caption{\label{Figure:CrossSectionsDeg}(Color online) Differential cross sections for $\gamma p\to 
    p\pi^0$ from CBELSA/TAPS ({\large\color{red}$\bullet$}) plotted versus $\theta_{\rm c.m.}$. For 
    comparison, CB-ELSA data~\cite{Bartholomy:2004uz} are represented by 
    ({\tiny\color{blue}$\blacksquare$}) and CLAS data~\cite{Dugger:2007bt} by 
    ({\large\color{cyan}$\circ$}). Note that the CLAS cross sections
    were extracted for 50-MeV-wide energy bins; for $E_\gamma < 1.3$~GeV, 
    the same CLAS results are thus shown in the corresponding
    25-MeV-wide bins chosen for this analysis. The solid blue line
    shows the solution of the Bonn-Gatchina
    partial-wave analysis~\cite{Anisovich:2010an}, and the dashed
    black line represents the current SAID solution (SP09) ~\cite{Dugger:2009pn}. 
    The data points include statistical errors only; the total
    systematic error is given as error bands at the bottom of each plot.}
\end{figure*}

The $P_\gamma$ error is assumed to dominate the total error of the photon flux, which has a 
strong dependence on the efficiency of the beam monitor at the end of the beam line 
(Fig.~\ref{Figure:Experiment}). By varying the background subtraction of noncoincident 
tagger-beam monitor hits, $P_\gamma$ was determined to be $0.639\,\pm\, 0.002\,_{\rm stat.}\,\pm\, 
0.05\,_{\rm sys.}$~\cite{Crede:2009zz}. This value agrees with determinations from multiple 
Tagger-Or-Runs at different incoming photon rates. An overall error of 10\,\% has been assigned
to the photon flux determination and has been included in Figs.~\ref{Figure:CrossSections} and
\ref{Figure:CrossSectionsDeg}.
\begin{figure*}[ht]
  \epsfig{file=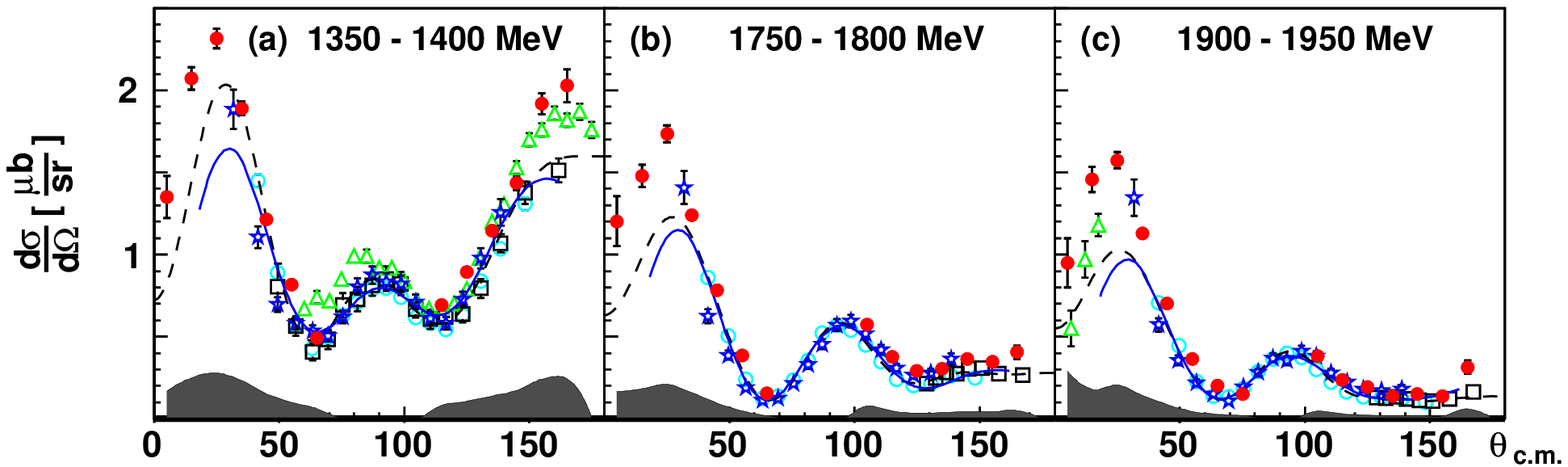,width=1.0\textwidth}
  \caption{\label{Figure:Selected}(Color online) The differential cross sections for $\gamma p\to
    p\pi^0$ at $E_\gamma\in [1350,1400]$~MeV (left), $E_\gamma\in [1750,1800]$~MeV (center), and
    $E_\gamma\in [1900,1950]$~MeV (right). The angle shown is the pion center-of-mass scattering
    angle. The experimental data are from the CBELSA/TAPS measurements presented here (filled 
    circles {\large\color{red}$\bullet$}), CB-ELSA~\cite{Bartholomy:2004uz}~(open stars 
    {\color{blue}\ding{73}}), GRAAL (left)~\cite{Bartalini:2005wx}~(open boxes {\scriptsize$\square$}), 
    SPring-8~(center and right)~\cite{Sumihama:2007qa}~(open boxes {\scriptsize$\square$}), 
    CLAS~\cite{Dugger:2007bt}~(open circles {\large\color{cyan}$\circ$}), Althoff et al.~(left)
    \cite{Althoff:1979mc}~(open triangles {\color{green}$\vartriangle$}), and Brefeld et al.~(right)
    \cite{Brefeld:1975dv}~(open triangles {\color{green}$\vartriangle$}). The dashed lines represent
    the current SAID solution (SP09)~\cite{Dugger:2009pn}, and the solid lines represent the latest 
    solution of the Bonn-Gatchina PWA~\cite{Anisovich:2010an}, which includes the new data 
    presented here. The total systematic error is given again as an error band at the bottom of each figure.}
\end{figure*}

\subsection{Systematic Uncertainties\label{Subsection:SystematicUncertainties}}
The statistical errors are determined from the number of events in each 
($E_\gamma$,\,cos\,$\theta_{\rm c.m.}$) or ($E_\gamma$,\,$\theta_{\rm c.m.}$) 
bin. Statistical errors are shown for all data points; systematic uncertainties 
are given as error bands at the bottom of each plot.

The systematic errors include uncertainties from the position of the liquid hydrogen target 
and a possible offset of the photon beam. The position of the target cell was determined 
from kinematic fitting by comparing the off-zero displacement of different pull distributions 
to Monte Carlo simulations. It was found to be shifted upstream by 0.65~cm~\cite{vanPee:2007tw}. 
The corresponding systematic errors were determined by varying the target position in the Monte 
Carlo $(\pm 1.5~{\rm mm})$ and evaluating changes in the reextracted differential cross sections. 
The errors show an angular dependence, but are 3-4\,\% on average and $< 7$\,\% at most in the 
very forward region, $\theta_{\rm c.m.} < 50^\circ$. The photon beam was assumed to be shifted by 
less than 2~mm off axis at the target position. The errors of the decay branching fractions are 
negligible.

The reconstruction of photons from the decay of neutral mesons and the
identification of final states requires a sequence of cuts including the use 
of kinematic fitting. In \cite{Crede:2009zz}, it was discussed in detail that
the reconstruction of $\eta$~mesons from final states with two and six
photons leads to compatible results for the $\eta$~differential cross
sections. The reconstruction of $\pi^0~(\to\gamma\gamma)$~mesons 
is very similar to  the reconstruction of $\eta$~mesons in the same two-photon 
decay mode and for the $\pi^0$~analysis discussed here, the same data sets,
software packages, and calibration were used. This shows the good
understanding of the detector response to multi-photon final
states. An overall $\pm\,5.7\,\%$ error is assigned to the
reconstruction efficiency as determined in~\cite{Amsler:1993kg}. An
additional 3\,\% systematic error accounts for the slightly different
effects of CL cuts on data and Monte Carlo events. The
uncertainty of the proton trigger has been estimated from the small
disagreement of the differential $\eta$~cross sections using the
$\eta\to 2\gamma$ and $\eta\to 3\pi^0\to 6\gamma$ decay channels for
$E_\gamma < 1$~GeV and cos\,$\theta_{\rm c.m.} < 0.0$ where the proton
trigger dominates over triggers caused by photons.

All the discussed contributions to the systematic error have been added
quadratically to give the total systematic uncertainty for this reaction.

\section{Experimental Results\label{Section:Results}}
Figure~\ref{Figure:CrossSections} shows the $\gamma p\to p\pi^0$
differential cross sections plotted versus cos\,$\theta_{\rm c.m.}$ of
the $\pi^0$, using cos\,$\theta_{\rm c.m.}$ bins of width 0.1. To
facilitate the comparison with our previous CB-ELSA data, we have
chosen energy bins of 25~MeV for the energy range
$E_\gamma\in[850,1300]$~MeV and energy bins of 50~MeV for the energy
range $E_\gamma\in [1300,2500]$~MeV. The data cover the very forward
region. We have excluded those data points in this analysis that show
a Monte Carlo acceptance of $< 5\%$. The error bars are statistical
only; the total systematic error is given as error bands at the bottom
of each plot. The angular distributions exhibit a region of zero acceptance 
around approximately $-0.3 < {\rm cos}\,\theta_{\rm c.m.} < 0.3$, 
depending on energy, because of the trigger conditions during the data
taking. We observed a small drop of the cross section for $E_\gamma\in
[1075,1100]$~MeV, owing to an increased photon flux determined for one 
of the tagger fibers defining this energy bin; the reason for this
increase is unknown, and thus these data are not shown. Resonance
production is clearly observed up to high energies, as can be seen
from variations in the differential cross sections over the full
angular range up to high energies. Above $E_\gamma > 1.5$~GeV, the
development of a forward peak indicates important contributions from
$t$-channel exchange, which becomes dominant above incoming energies
of 2.2~GeV.

Figure~\ref{Figure:CrossSectionsDeg} shows the same $\gamma p\to
p\pi^0$ differential cross sections plotted versus $\theta_{\rm c.m.}$
of the $\pi^0$ to illustrate the improvement of the data coverage
in the forward direction. The width of each $\theta_{\rm c.m.}$~bin is
$10^\circ$. This change of the data representation and the modified
binning required a new analysis and subtraction of the background 
under the $\pi^0$~peaks in the mass spectra as well as a new
investigation of the systematic errors. Again, the total systematic
error is given as error bands at the bottom of each plot. For this
representation of the data, we have excluded data points that show a
Monte Carlo acceptance of $< 1\%$. Figure~\ref{Figure:Selected} shows
the differential cross sections at three different incoming photon
energies to facilitate the comparison with a variety of other data
sets.

\subsection{Comparison with CB-ELSA and CLAS Data\label{Subsection:Comparison}}
\begin{figure}[b]
 \epsfig{file=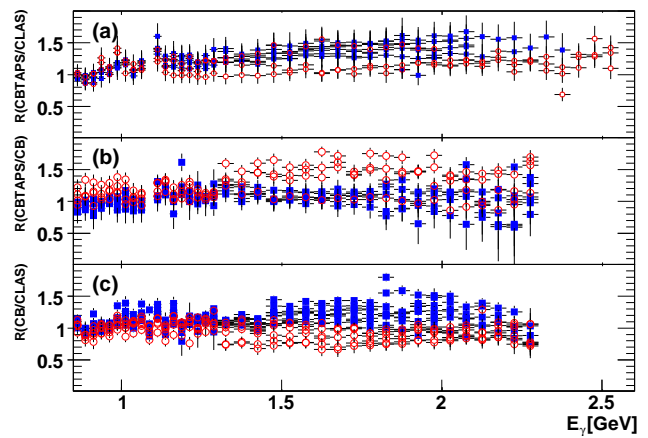,width=0.49\textwidth}
  \caption{\label{Figure:Ratios}(Color online) Ratios of differential
    cross sections from this analysis to CLAS~\cite{Dugger:2007bt}~(A)
    and to CB-ELSA~\cite{Bartholomy:2004uz}~(B), as well as from
    CB-ELSA to CLAS~(C). The red open circles represent the forward
    direction $({\rm cos}\,\theta_{\pi^0} > 0)$ and the blue squares the
    backward direction $({\rm cos}\,\theta_{\pi^0} < 0)$. For
    $E_\gamma < 1.3$~GeV, the same CLAS results have been used
    for the corresponding 25-MeV-wide bins chosen for this analysis.}
\end{figure}

A more detailed comparison of the results in the form of cross section
ratios of this analysis to other data sets is shown in Fig.~\ref{Figure:Ratios}. 
The red open circles represent the forward direction, ${\rm cos}\,\theta_{\pi^0} 
> 0$, and the solid blue squares the backward direction, ${\rm cos}\,\theta_{\pi^0} 
< 0$. In Fig.~\ref{Figure:Ratios}\,(a), we compare the new CBELSA/TAPS data
of this analysis with the previous CLAS data~\cite{Dugger:2007bt} and in 
Fig.~\ref{Figure:Ratios}\,(b), with the previous CB-ELSA data~\cite{Bartholomy:2004uz}. 
In Fig.~\ref{Figure:Ratios}\,(c), a comparison between CB-ELSA and CLAS
is shown. Since the earlier CB-ELSA data switch from a 50-MeV to a 100-MeV 
energy binning at high energies owing to statistical limitations, the
distributions in panels (b) and (c) are cut off at an incoming photon energy of 
2.3~GeV. 

\begin{figure*}[ht]
 \epsfig{file=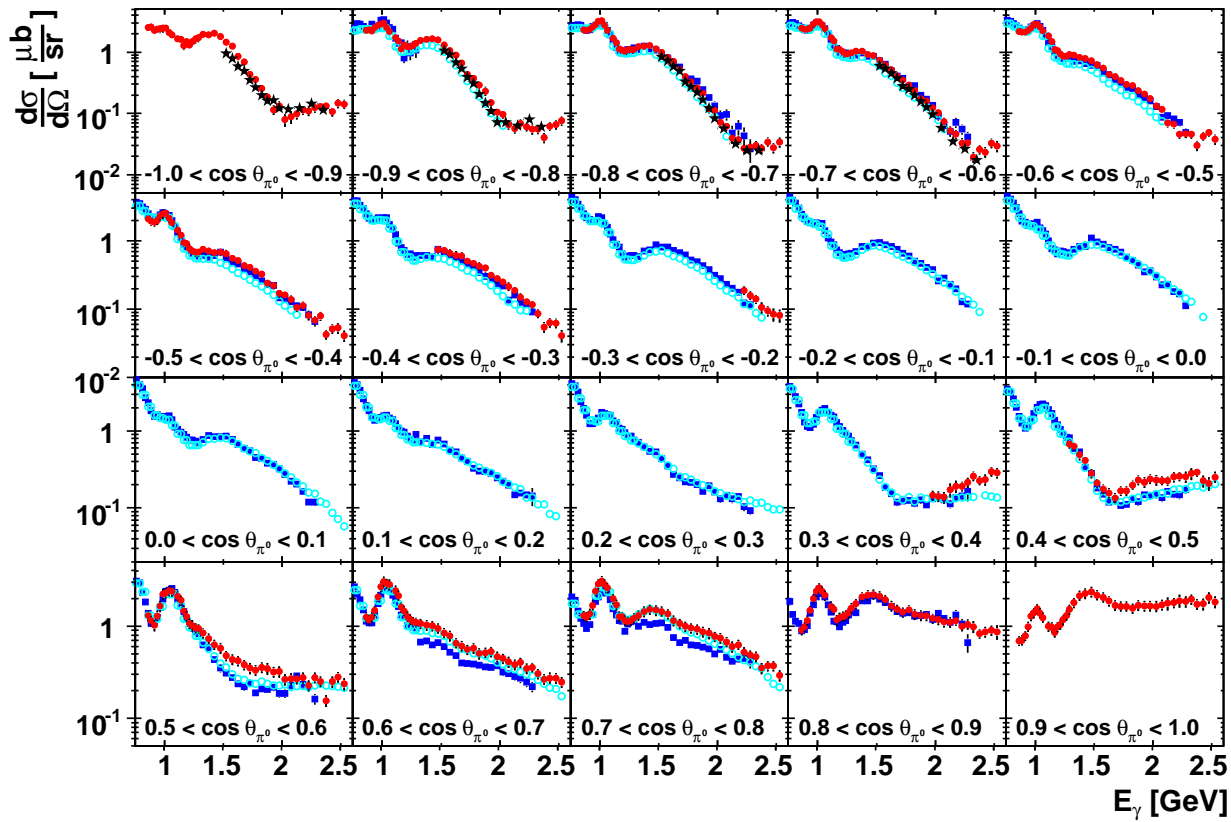,width=0.95\textwidth}
  \caption{\label{Figure:Comparison}(Color online) Fixed-angle
    excitation functions for the reaction $\gamma p\to p\pi^0$; the
    angle shown is the pion center-of-mass scattering angle. The
    experimental data are from the new analysis presented here
    (filled circles {\large\color{red}$\bullet$}), CB-ELSA~\cite{Bartholomy:2004uz}
    ({\tiny\color{blue}$\blacksquare$}), CLAS~\cite{Dugger:2007bt}~(open 
    circles {\large\color{cyan}$\circ$}), and LEPS~\cite{Sumihama:2007qa} 
    (filled {\large $\star$}). Note that the CLAS cross sections were extracted for 
    50-MeV-wide energy bins; for $E_\gamma < 1.3$~GeV, the same CLAS
    results are shown in the corresponding 25-MeV-wide bins chosen
    for this analysis.}
\end{figure*}

Figure~\ref{Figure:Comparison} shows the excitation functions
for the reaction $\gamma p\to p\pi^0$ at fixed values of the pion
center-of-mass scattering angle using a logarithmic scale. We also
added available data points in the backward direction from 
LEPS~\cite{Sumihama:2007qa}. The agreement with LEPS is good and
confirms the weak energy dependence of the cross section found above
2~GeV and at very backward angles. A very similar, mostly flat distribution
is observed at very forward angles.

In a direct comparison, our new data show an overall good agreement
with the older CB-ELSA data within systematic errors
[Fig.~\ref{Figure:Ratios}\,(b)]. While our new data show a steeper slope
of the forward ($t$-channel) rise than the earlier CB-ELSA results,
the most forward CB-ELSA point at $\theta_{\rm c.m.}\approx 35^\circ$
is almost perfectly matched for all energies, indicating the very good
agreement of the overall strength of the forward
peak~(Fig.~\ref{Figure:Comparison}). The few red points in
Fig.~\ref{Figure:Ratios}\,(b), which are systematically higher between
1.5~and 2.3~GeV, indicate the different slope observed in this analysis. 
A less rapid increase of the differential cross section in this kinematic 
region, $0.3 < {\rm cos}\,\theta_{\pi^0}$, which was implied by the 
recent CLAS data and suggested in ~\cite{Dugger:2007bt}, can thus not
be confirmed in this analysis.


Below $E_\gamma < 1.3$~GeV, the overall discrepancies between CBELSA/TAPS
and CLAS are smaller than about 20\,\%, and thus, both data sets are
consistent within systematic errors [Fig.~\ref{Figure:Ratios}\,(a)]. At
higher energies, the forward direction agrees fairly well between the
two data sets and appears to be consistent within systematic errors. 
Only a somewhat higher band of forward points indicates again the 
steeper slope of the forward peak observed in this analysis. However,
agreement with CLAS for $0.6 < {\rm cos}\,\theta_{\pi^0}$ is fair to
good (Fig.~\ref{Figure:Comparison}). Note that additional systematic
normalization errors of the CLAS and CB-ELSA data (discussed 
in~\cite{Dugger:2007bt,vanPee:2007tw}, but not included in the published 
data) are not considered in this comparison. Larger systematic differences 
up to 50\,\% at and above $E_\gamma\approx 1.5$~GeV remain only in 
the backward direction. This confirms the findings of the previous 
CB-ELSA analysis, which also observed a higher differential cross section 
in the backward region relative to the CLAS data [Fig.~\ref{Figure:Ratios}\,(c)]. 
Older data from Bonn~\cite{Althoff:1979mc} also support these higher 
cross sections~[Fig.~\ref{Figure:Selected}\,(a)]. 

This forward-backward disagreement suggests a non-trivial scale
discrepancy between CLAS and CBELSA/TAPS, and we conclude that a
normalization effect can be ruled out. Most of the recent data sets
are not statistically limited, and a reason for the observed differences 
is not known. Since all the discussed discrepancies show the overall
level of uncertainty for the reaction $\gamma p\to p\pi^0$ and this
has a significant impact on the extraction of resonance couplings,
this issue needs to be further addressed and resolved by future
experiments.

\section{\label{Section:PWA} Partial-Wave Analysis}
A partial-wave analysis (PWA) has been performed within the
Bonn-Gatchina PWA framework and includes the data presented 
here. For the discussion of contributing resonances, we use the
conventional naming scheme of the PDG~\cite{Nakamura:2010zzi}:
\begin{equation}
N({\rm mass~of~the~resonance})\,L\,_{2I,\,2J}\,,
\end{equation}
where $I$ and $J$ denote isospin and total spin of the resonance,
respectively, and $L$ the orbital angular momentum in the decay into
a nucleon and a pion. The scattering amplitudes in the PWA for the
production and decay of baryon resonances are constructed in the
framework of the spin-momentum operator expansion method
suggested in~\cite{Anisovich:2001ra,Anisovich:2004zz,Anisovich:2006bc}. 
The approach is relativistically invariant and allows one to perform
combined analyses of different reactions imposing analyticity and
unitarity directly. The analysis method is further documented
in~\cite{Klempt:2006sa,Anisovich:2007zz}. The experimental 
database used for the PWA is described in detail
in~\cite{Anisovich:2010an} and includes the number of data points for
each reaction as well as the weight for each data set used in the fits.
It takes into account almost all important sets of photo- and pion-induced 
reactions, including three-body final states. The latter were included 
in event-based likelihood fits for the PWA solution discussed 
in~\cite{Anisovich:2010an}. A full description of the experimental
database goes beyond the scope of this paper, and thus we limit 
the discussion to $\pi^0$~photoproduction data. 

\begin{figure}[t]
\epsfig{file=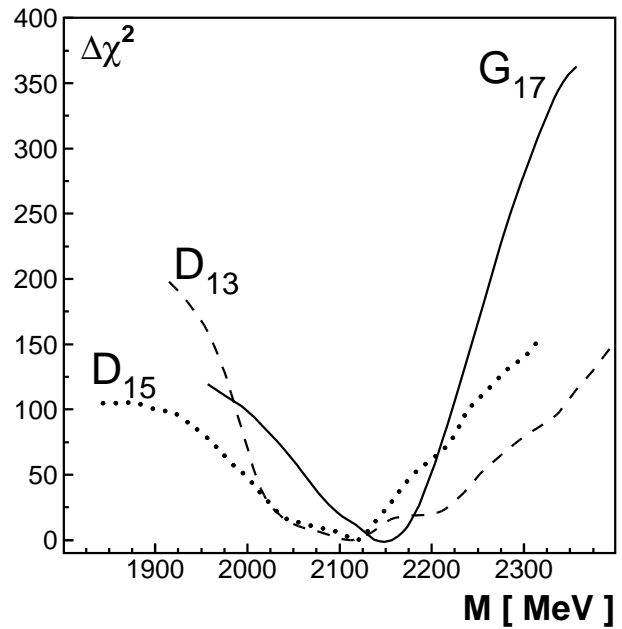,width=0.47\textwidth}
\caption{\label{Figure:MassScan} Mass scans of the $G_{17}$~state (solid
  line), the second $D_{15}$~state (dotted line), and the fourth 
  $D_{13}$~state (dashed line), which show the deviations in fit quality 
  from the optimum fit ($\Delta \chi^2$) for a series of mass values.}
\end{figure}

For the $\gamma p\to p\pi^0$ differential cross sections, data for 
the invariant mass region above 1600~MeV/$c^2$ are available from
GRAAL~\cite{Bartalini:2005wx}, \text{CB-ELSA}~\cite{Bartholomy:2004uz,
vanPee:2007tw}, and CLAS~\cite{Dugger:2007bt}. However, the 
GRAAL data are available only up to 1950~MeV/$c^2$ and the CLAS 
data cover only a rather restricted angular range at large masses. 
The new CBELSA/TAPS data cover the forward and backward angular 
region up to 2370~MeV/$c^2$ and therefore provide important
information about states in the third and fourth 
resonance regions. In particular, the new data are rather sensitive 
to contributions from high-spin resonances located in the mass 
region $2050-2150$~MeV/$c^2$. The behavior of the differential 
cross sections at very extreme angles should be a good indicator for 
the presence of such states. The sensitivity of the PWA solution to
a particular resonance is studied by omitting or including it in 
the fits and evaluating the fit quality in terms of the resulting 
$\chi^2$~values. The observation of a clear minimum in a mass scan 
gives further confidence about a contributing resonance. In such 
a scan, the mass of a resonance is chosen in several steps in the 
vicinity of the optimum value and fixed in the fit. The mass 
is then plotted versus the change of the $\chi^2$~value from the 
best fit. Figure~\ref{Figure:MassScan} shows a typical scan from 
this analysis. 

%The new data are fully compatible in our analysis with the previous 
%measurements of the CB-ELSA collaboration~\cite{vanPee:2007tw} 
%if the data points are scaled down by 14\%. 
An overall scaling factor of 14\,\% is needed in the PWA to achieve 
quantitative consistency of the data with the previous measurements 
of the CB-ELSA collaboration~\cite{vanPee:2007tw}. It is worth noting 
that a similar factor was needed to provide a compatibility between the 
CB-ELSA and CLAS data~\cite{Dugger:2007bt}. In that case, however, the 
CLAS data needed to be scaled up by 5\,\%. Fitting only such a scale 
factor, we obtained a rather good description of the data (reduced 
$\chi^2=1.46$) for both solutions reported in~\cite{Anisovich:2010an}. 
Including the new data in the database and slightly readjusting the 
parameters, we obtained an improved value of $\chi^2=1.26$ 
without notable changes in the description of other photoproduction 
data.

The highest spin state which was observed in photoproduction
reactions is the $N(2190)G_{17}$ resonance. This is a four-star
resonance according to the PDG classification, and the behavior of 
the $\pi N$~elastic amplitude in this partial wave clearly shows the
presence of a pole in the mass region $2100-2200$~MeV/$c^2$. The recent 
CLAS data on the reaction $\gamma p\to K\Lambda$~\cite{McCracken:2009ra} 
demand a contribution from this state for a good description of the 
recoil asymmetry data. Thus, we expect to observe this state in the 
pion photoproduction reactions. Indeed, we found that the coupling of 
this state into the $\pi N$~channel (which is fixed by the elastic data) 
is much stronger than that into the $K\Lambda$~channel, and in the 
combined analysis of our database, we found a small but not negligible 
contribution from this state to the pion photoproduction cross section. 
The pole position of the $G_{17}$~state was found to be $2170-i\,155$~MeV, 
and the mass scan %, describing deviations in fit quality from the optimum 
%fit ($\Delta \chi^2$) as a function of mass, 
is shown in Fig.~\ref{Figure:MassScan} with a solid line. A minimum, 
fully compatible with that observed in the CLAS $\gamma p\to K\Lambda$ 
data is clearly seen.

One of the new states which was observed in the earlier analysis of
the pion and $\eta$~photoproduction data~\cite{Anisovich:2005tf} was
a $D_{13}$~state with a mass of $2166^{+25}_{-50}$~MeV/$c^2$ and a
width of $300\pm65$~MeV/$c^2$. In the combined analysis of the present
database, this state was optimized with a mass of $2130\pm 20$~MeV/$c^2$ 
and a width of $310\pm 30$~MeV/$c^2$. The mass scan of this state for the 
presented pion photoproduction data is shown in Fig.~\ref{Figure:MassScan} 
as a dashed line. It is seen that the minimum corresponds well to the mass 
of the state found in the combined analysis, but the data are more tolerant 
to a lower resonance mass.

For the first observation of the $N(2070)D_{15}$~state in the analysis 
of the $\gamma p\to p\eta$ differential cross section~\cite{Anisovich:2005tf}, 
it was shown that this state can also contribute to the pion photoproduction 
cross section. In the present analysis, the $\pi N$~coupling of this state 
was found to be similar to the $\eta N$~coupling. The mass scan, shown in
Fig.~\ref{Figure:MassScan} with a dotted line, produced the minimum slightly 
above the mass value defined by the combined fit. However, the data
can accept well the mass of the state between 2060 and 2120~MeV/$c^2$.

Higher-mass $F$-wave states can also contribute to photoproduction and 
were included as multichannel relativistic Breit-Wigner amplitudes
in the PWA~\cite{Anisovich:2010an}. We have performed 
mass scans of the $N(2000)F_{15}$, $N(1990)F_{17}$, and $N(1890)S_{11}$~states, 
which were observed in the analysis of the $\gamma p\to K\Lambda$ CLAS 
data~\cite{McCracken:2009ra}. Candidates for these $F$-wave states are
listed as two-star resonances by the PDG; a third $S_{11}$ resonance has 
been assigned a one-star status~\cite{Nakamura:2010zzi}. Evidence for 
a higher-mass $F_{15}$-state around 2000~MeV/$c^2$ was also recently 
reported by the CLAS collaboration in $\omega$~photoproduction off the 
proton~\cite{:2009rc}. However, no minima were observed in our mass
scans for the description of the present data. Further single- and
double-polarization observables are needed to constrain the scattering
amplitude and to extract weakly contributing resonances.

In a final comment, we note that our description of the data and the
current SAID solution (SP09)~\cite{Dugger:2009pn} underestimate 
the most-forward data at all three photon energies presented in
Fig.~\ref{Figure:Selected}. The overall systematic error can certainly
account for some of the observed discrepancy, but some physics may be
missing. This needs to be further addressed in future analyses.

\section{\label{Section:Summary}Summary}
In summary, we have presented data on the photoproduced $\pi^0$~cross 
section in the reaction $\gamma p\to p\pi^0$. The continuous beam from 
the ELSA accelerator and the fiber detector of the tagging system provided 
tagged photons in the energy range from 850 to 2500~MeV. The results are in 
good overall agreement with previous measurements of this reaction but
are underestimated by the models at very forward angles for large photon 
energies. The Bonn-Gatchina PWA found the $N(2190)G_{17}$, $N(2080)D_{13}$, 
and $N(2070)D_{15}$ high-mass resonances in its description of the data.

\subsection*{Acknowledgments}
We thank the technical staff at ELSA and at all the participating institutions 
for their invaluable contributions to the success of the experiment. We 
acknowledge financial support from the National Science Foundation (NSF), 
Deutsche Forschungsgemeinschaft (DFG) within the SFB/TR16 and from 
Schweizerischer Nationalfonds. The collaboration with St. Petersburg 
received funds from DFG and the Russian Foundation for Basic Research.

\end{document}